\documentclass[10pt]{iopart}
\usepackage{citesort}
\usepackage[utf8]{inputenc} 
\usepackage{graphicx}
\usepackage{amsfonts}
\newcommand{\be}{\begin{equation}}
\newcommand{\ee}{\end{equation}}
\newcommand{\bea}{\begin{eqnarray}}
\newcommand{\eea}{\end{eqnarray}}

\usepackage[margin=1.25in]{geometry}

\usepackage{color}

\begin{document}
\title{Bose-Einstein condensation of $^{162}$Dy and $^{160}$Dy}
\author{Yijun Tang, Nathaniel Q. Burdick, Kristian Baumann, and \\Benjamin L. Lev}
\address{Department of Applied Physics, Stanford University, Stanford CA 94305, USA}
\address{Department of Physics, Stanford University, Stanford CA 94305, USA}
\address{E. L. Ginzton Laboratory, Stanford University, Stanford CA 94305, USA}
\ead{benlev@stanford.edu}

\date{\today}
\begin{abstract}
We report Bose-Einstein condensation of two isotopes of the highly magnetic element dysprosium: $^{162}$Dy and ${^{160}}$Dy. For $^{162}$Dy, condensates with  $10^5$ atoms form below $T=50$~nK. We find the evaporation efficiency for the isotope $^{160}$Dy to be poor; however, by utilizing a low-field Fano-Feshbach resonance to carefully change the scattering properties, it is possible to produce a  Bose-Einstein condensate of $^{160}$Dy with $10^3$ atoms. The $^{162}$Dy BEC reported is an order of magnitude larger in atom number than that of the previously reported $^{164}$Dy BEC, and it may be produced within 18 s.

\end{abstract}

\section{Introduction}
Dipolar quantum gases expand the scope of ultracold atom experiments by introducing  anisotropic and long-range interactions, providing new avenues for exploring correlated many-body systems~\cite{Lahaye09}. Experiments employing the open-shell lanthanides---and in particular dysprosium (Dy), an element with unsurpassed magnetic moment---offer the possibility to extend the foundational experiments performed with chromium ($6\ \mu_{B}$)~\cite{CrBEC} to new regimes of ultracold dipolar physics.   In particular, bosonic and fermionic quantum degenerate gases have been produced with Dy   ($10\ \mu_{B}$)~\cite{Dy164,Dy161} and erbium ($7\ \mu_{B}$) \cite{ErBEC,ErDFG}.    Specifically, of Dy's five naturally abundant isotopes---two fermions, three bosons---quantum gases of fermionic $^{161}$Dy~\cite{Dy164} and  bosonic  $^{164}$Dy~\cite{Dy161} have been produced. We report Bose-Einstein condensates (BECs) of the two remaining high-abundance bosonic isotopes, $^{162}$Dy and $^{160}$Dy.   Notably, we  find $^{162}$Dy  condensates   are ten-times larger than those of $^{164}$Dy, and may be produced in less than 18 s, while $^{160}$Dy are tiny and may only be condensed through careful use of a low-field Fano-Feshbach resonance~\cite{FBPRA}.

\section{Initial loading and state preparation}

Initial laser cooling and trapping of Dy atoms is performed using techniques explained in previous works \cite{Dy164,FBPRA}. In brief, atoms from a high-temperature effusive cell are loaded into a magneto-optical trap (MOT) via a Zeeman slower, both with 421-nm laser light. The loading time for this MOT stage is 5~s for $^{162}$Dy and 10~s for $^{160}$Dy. This difference is due to their different natural abundance: 25.5\% for $^{162}$Dy and 2.3\% for $^{160}$Dy. The atoms are subsequently transferred into a blue-detuned, narrow-linewidth MOT at 741~nm for 5~s, reaching a final atom number of $4\times10^7$ for $^{162}$Dy and $7\times10^6$ for $^{160}$Dy, both at $T\approx 2\ \mathrm{\mu K}$ and spin-polarized in $|J=8,m_J=+8\rangle$.  The atomic cloud is subsequently loaded into a horizontal optical dipole trap (ODT1) (see figure~\ref{Setup}) formed by a 1064-nm laser beam with waists of $24(2)\ \mu\mathrm{m}$ and $22(2)\ \mu\mathrm{m}$ and an initial power of 5.0(3) W. To improve loading efficiency, we broaden the trap to an aspect ratio of $\sim$5 by scanning the beam horizontally with an acousto-optic modulator~\cite{ErThesis}. Once loaded into ODT1, the atoms are transferred to the absolute electronic ground state $|J=8,m_J=-8\rangle$ by radio-frequency-induced adiabatic rapid passage. Loading into this state eliminates loss due to rapid dipolar relaxation~\cite{Burdick:2014va}.  After compressing this trap by ramping off the horizontal modulation and increasing the power, we have $\sim$$5\times10^6$ $^{162}$Dy or $\sim$$2\times10^6$ $^{160}$Dy atoms at a temperature of 5-10 $\mu$K, which sets the starting point for forced evaporative cooling.

\section{Bose-Einstein condensation of $^{162}$Dy}

\begin{figure}[t]
\includegraphics[width=0.6\columnwidth]{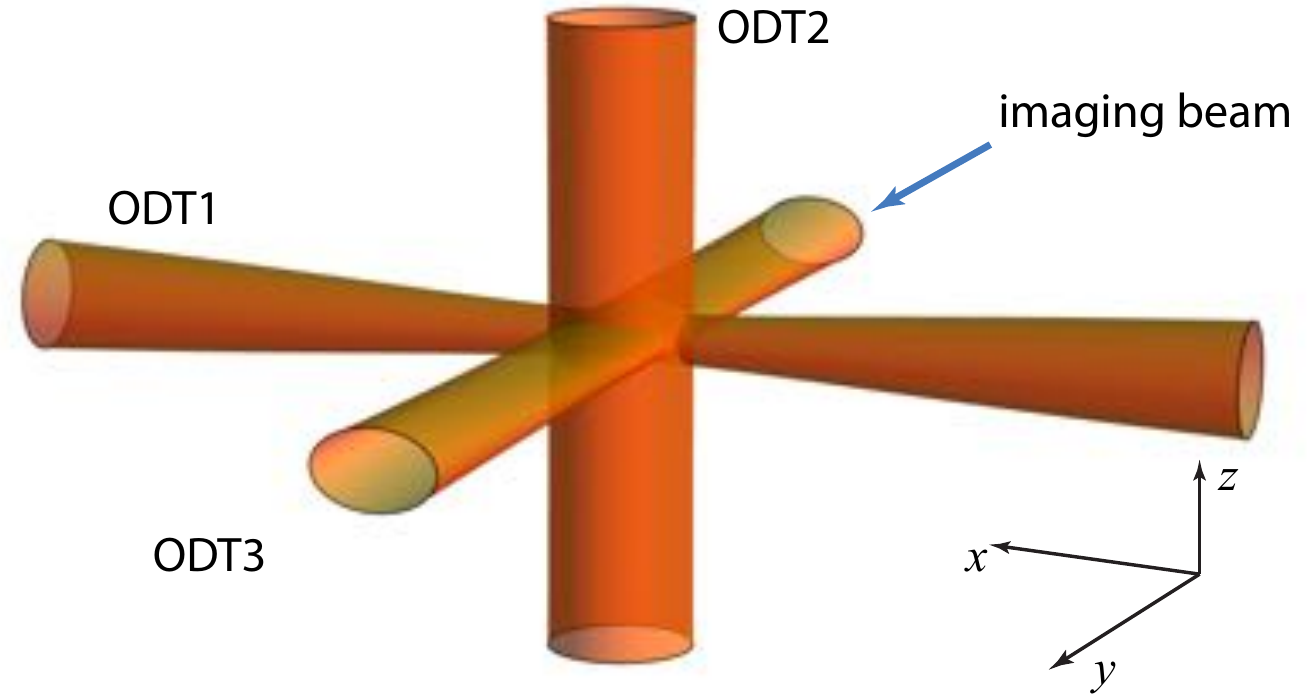}
\caption{Orientation of optical dipole traps and imaging axis in experimental setup.} 
\label{Setup}
\end{figure}

We perform forced evaporative cooling of $^{162}$Dy at a constant vertical magnetic field $B_z=1.568(5)$~G to avoid  Fano-Feshbach resonances~\cite{FBPRA}. Our evaporative scheme employs two crossed optical dipole traps (cODTs) consisting of three 1064-nm beams shown in figure \ref{Setup}. The beams in both cODTs are detuned by $\sim$160~MHz to avoid interferences.

The first cODT is formed by crossing  ODT1 with a vertical beam (ODT2) of  circular waist  $75(2)\ \mu\mathrm{m}$. Evaporation begins by reducing the powers of ODT1 and ODT2 for 2~s from 5.0(3)~W to 1.0(1)~W and 4.0(2)~W to 3.2(2)~W, respectively. Both ramps take the form $P(t)=P_i/(1+t/\tau)^\beta$, where $t$ is time, and $\tau$ and $\beta$ are experimentally optimized parameters \cite{ODTRamp}. The phase space density of the atomic cloud is two orders of magnitude away from condensation at this stage. 

We load the atoms into the second cODT in 1~s by linearly increasing the power of ODT2 to 5.6(3)~W, linearly reducing the power of ODT1 to 0~W, and by ramping up a second horizontal 1064-nm beam (ODT3) to 2.5(1)~W. ODT3 is cylindrical with a horizontal waist of $65(2)\ \mu\mathrm{m}$ and vertical waist of $35(2)\ \mu\mathrm{m}$. While the first cODT is very tight for efficient initial evaporation, the second cODT is larger to avoid inelastic three-body collisions.  This second cODT is oblate, with the tight axis along $\hat{z}$, to avoid trap instabilities due to the dipole-dipole interaction (DDI)~\cite{CrNature,Dy164}.

\begin{figure}[t]
\includegraphics[width=0.9\columnwidth]{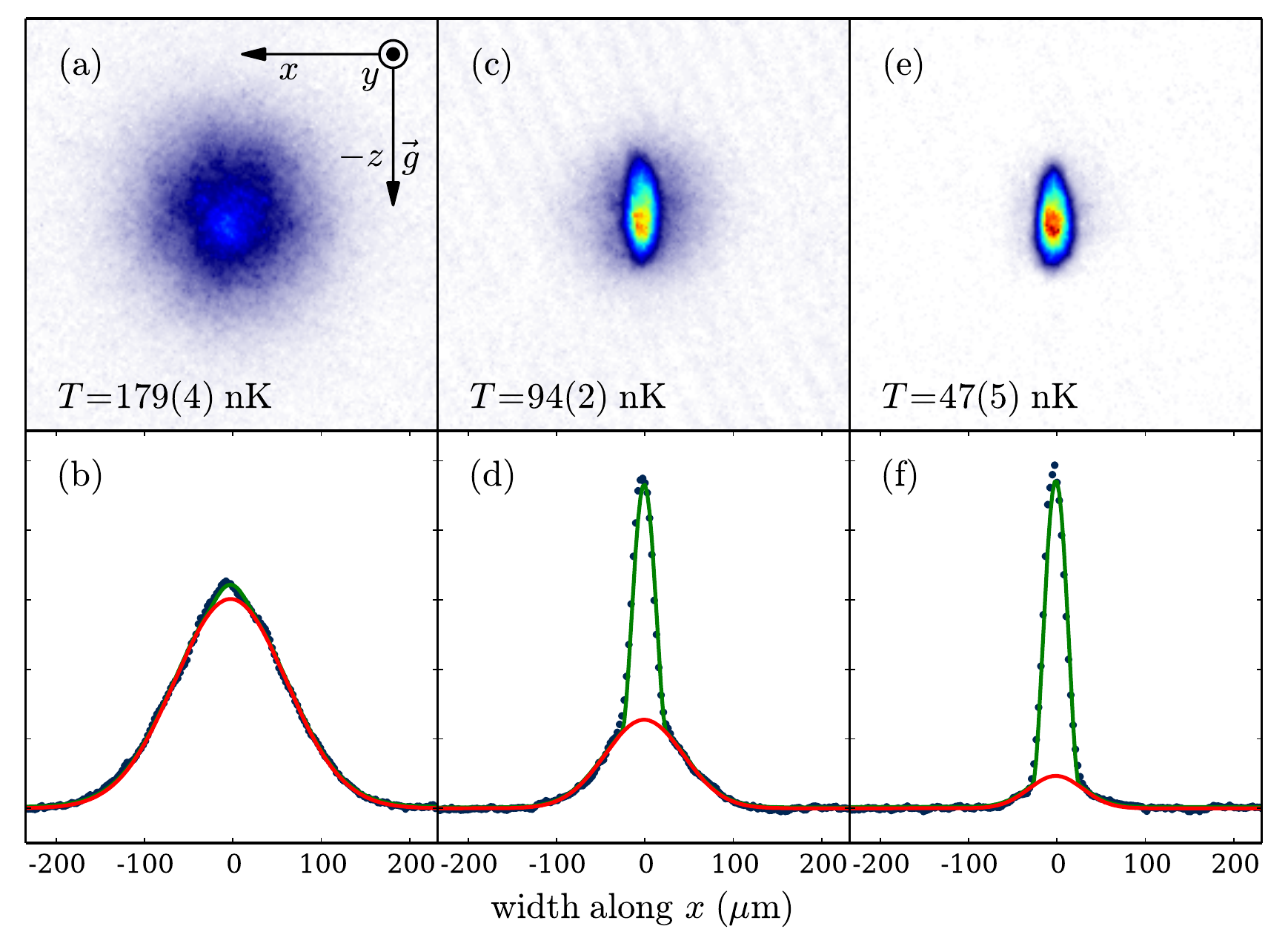}
\caption{Typical single-shot density profiles of $^{162}$Dy atomic clouds after 20 ms of free expansion. The top row shows the density distribution, with the colorscale equal for the three pictures. The lower row shows the integrated density along $\hat{x}$. The dots represent the data, the green curve shows the fitted bimodal profile and the red curve shows the fitted thermal portion. (a) and (b): $3.0(2)\times10^5$ atoms at the onset of BEC. $T=179(4)$ nK and BEC fraction is 3\%. (c) and (d): Intermediate stage of BEC with $1.5(2)\times10^5$ atoms at $T=94(2)$ nK. BEC fraction is 40\%. (e) and (f): Nearly pure BEC of $9.6(4)\times10^4$ $^{162}$Dy atoms at $T=47(5)$ nK. BEC fraction is 76\%.} 
\label{162BEC}
\end{figure}

The second evaporation stage begins with $\sim5\times10^5$ atoms. The powers of ODT2 and ODT3 are reduced in 5~s according to the same functional form provided above. Beginning with loading the first MOT stage, the total time to a pure $^{162}$Dy BEC is 18~s. After evaporative cooling, the atomic cloud is released for imaging by quickly ($<$10~$\mu$s) turning off the trapping beams, whereby the cloud expands freely for a variable time-of-flight (TOF). At the end of this TOF, we perform absorption imaging using resonant 421-nm light, where the magnetic field is kept constant along $\hat{z}$ until rapidly switching to $\hat{y}$~1~ms before imaging.  This provides a quantization axis for the $\sigma^-$-transition, which has the largest photon scattering cross section for the $|J=8,m_J=-8\rangle$ state. Single-shot absorption images after 20-ms TOF are shown in figure \ref{162BEC}(a), (c), and (e), with their corresponding 1D integrated densities along $x$ shown in~\ref{162BEC}(b), (d), and (f). We fit the data with a bimodal  distribution consisting of a Gaussian and Thomas-Fermi profile to extract the temperature from the residual thermal contribution~\cite{Ketterle99,TOFTemp}.

Figure \ref{162BEC} shows the evolution of the atomic density as we cool toward quantum degeneracy. The panels a-b show the density very close to the transition point, where we measure $3.0(2)\times10^5$ atoms at $T=179(4)$ nK. As we lower the final trap depth, the density in the center of the cloud dramatically increases  while the thermal fraction decreases, as shown in panels c-d for $1.5(2)\times10^5$ atoms at $T=94(2)$ nK.  A nearly pure BEC of $9.6(4)\times10^4$ $^{162}$Dy atoms at $T=47(5)$~nK is shown in figure~\ref{162BEC} e-f. The thermal background is hardly visible, and the cloud has an aspect ratio of 2.6.   The evaporation is quite efficient, with the log of the ratio of phase-space density increase to atom loss exceeding three for most of the sequence.  The final powers of ODT2 and ODT3 are 400(20) mW and 310(20) mW, respectively. We measured the final trap frequencies to be [$\omega_x,\omega_y,\omega_z$] = $2\pi\times$[66(2), 30(3), 131(1)] Hz, resulting in an oblate-trap aspect ratio of $\lambda=\omega_z/\sqrt{\omega_x\omega_y}\approx3$. If we assume $^{162}$Dy has a scattering length $a=100a_0$, where $a_0$ is the Bohr radius, the BEC has a peak density of $n_0\approx3\times10^{14}$ cm$^{-3}$ in the Thomas-Fermi limit.  The atom number and density are likely slightly larger than these reported values due to the optical saturation suffered while imaging at large optical depth. The BEC has a $1/e$ lifetime of 4.7(1)~s, shorter than our background lifetime of 21(1)~s.  Three-body losses will be discussed in a subsequent publication~\cite{BohnWithUs}.

\begin{figure}[t]
\includegraphics[width=0.8\columnwidth]{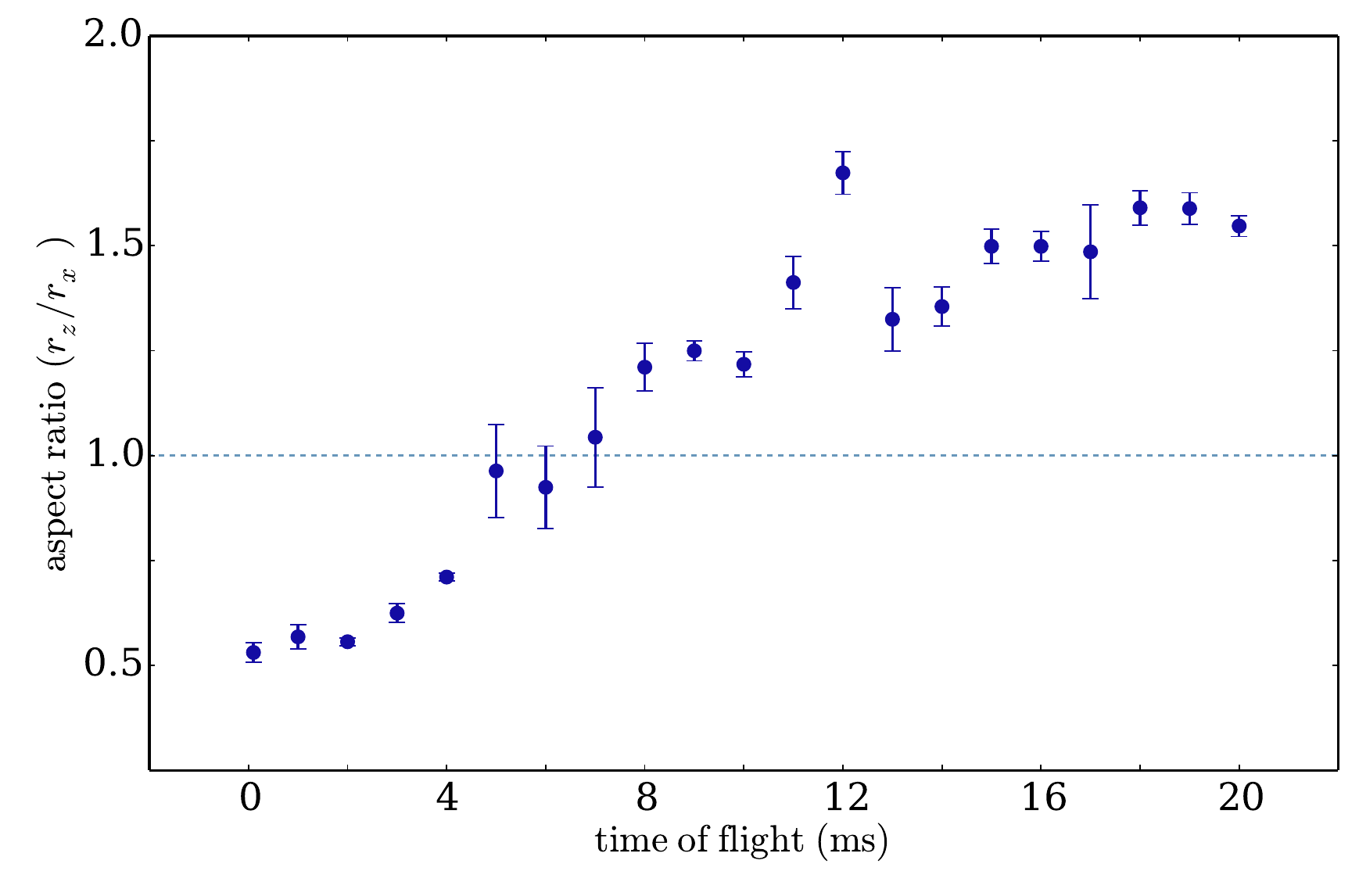}
\caption{Aspect ratio of the atomic density of a nearly pure BEC of $^{162}$Dy versus free expansion time. See text for details.} 
\label{162inversion}
\end{figure}

We measure the evolution of the cloud's aspect ratio during free expansion in further support of the observation of BEC. A thermal gas will expand into a spherical cloud at long TOF while a BEC will expand anisotropically, inverting its aspect ratio. The result of our measurement is shown in figure~\ref{162inversion}. The aspect ratio of the BEC in the trap is  $\sim$0.5, consistent with the measured trap frequencies. The aspect ratio is unity at  $\sim$6~ms TOF and reverses at longer TOFs, reaching 1.5 after 20 ms of TOF. As mentioned above, our bias field during evaporative cooling is in $\hat{z}$, whereas our imaging scheme requires a quantization field along $\hat{y}$. To measure the aspect ratio of the BEC without switching field direction during its evolution, we adiabatically rotate the field to $\hat{y}$  in 10 ms after the evaporation ramps.  We make sure that no  Fano-Feshbach resonances are encountered by maintaining a constant field magnitude, verified through rf spectroscopy, and we hold this configuration for 100 ms to ensure a steady state before releasing the cloud. The magnetic field is then kept constant during TOF and imaging. It is worth noting that for this field orientation, the final aspect ratio, 1.5, is not as large as the aspect ratio of 2.6 shown in figure \ref{162BEC}(e). This  dependence of aspect ratio on field angle is a manifestation of the anisotropy of the DDI~\cite{Lahaye09}.

For the measurement shown in figure~\ref{162inversion}, we drive $\sigma^+$ transitions for absorption imaging to avoid CCD saturation at short expansion times, where the optical density of the cloud is very high. This technique is based on the fact that for a high-spin atom like Dy, the transition strength in the stretched state is extremely different for $\sigma^-$ and $\sigma^+$ transitions. For bosonic Dy in the $|J=8,m_J=-8\rangle$ state, the transition strength ratio is $154:1$ \cite{LuThesis}. Hence imaging with resonant $\sigma^+$ light greatly reduces the effective optical density and is free of dispersive effects associated with off-resonant light. Because optical pumping occurs during imaging with $\sigma^+$ light, extracting the atom number from these images requires delicate calibration and detailed theoretical modeling. However,  aspect-ratio measurements are  largely unaffected by pumping effects during the 20-$\mu$s imaging pulse.

\section{Bose-Einstein condensation of $^{160}$Dy}

We use a very similar two-stage cooling scheme for the isotope $^{160}$Dy. The first stage of forced evaporative cooling is 3-s long, during which we ramp down the power of ODT1 from 5.0(3)~W to 1.3(1)~W and that of ODT2 from 5.6(3)~W to 2.0(1)~W. The initial power of ODT2 for the second stage is 5.6(3)~W and that of ODT3 is 3.0(2)~W. We found that evaporation efficiency is very low for $^{160}$Dy at fields free from the effects of Fano-Feshbach resonances, preventing the formation of BEC. This observation suggests that $^{160}$Dy has unfavorable scattering properties.  Publication of measurements of its scattering length, as well as those of $^{162}$Dy and $^{164}$Dy, are forthcoming~\cite{BohnWithUs}.

\begin{figure}[t]
\includegraphics[width=0.9\columnwidth]{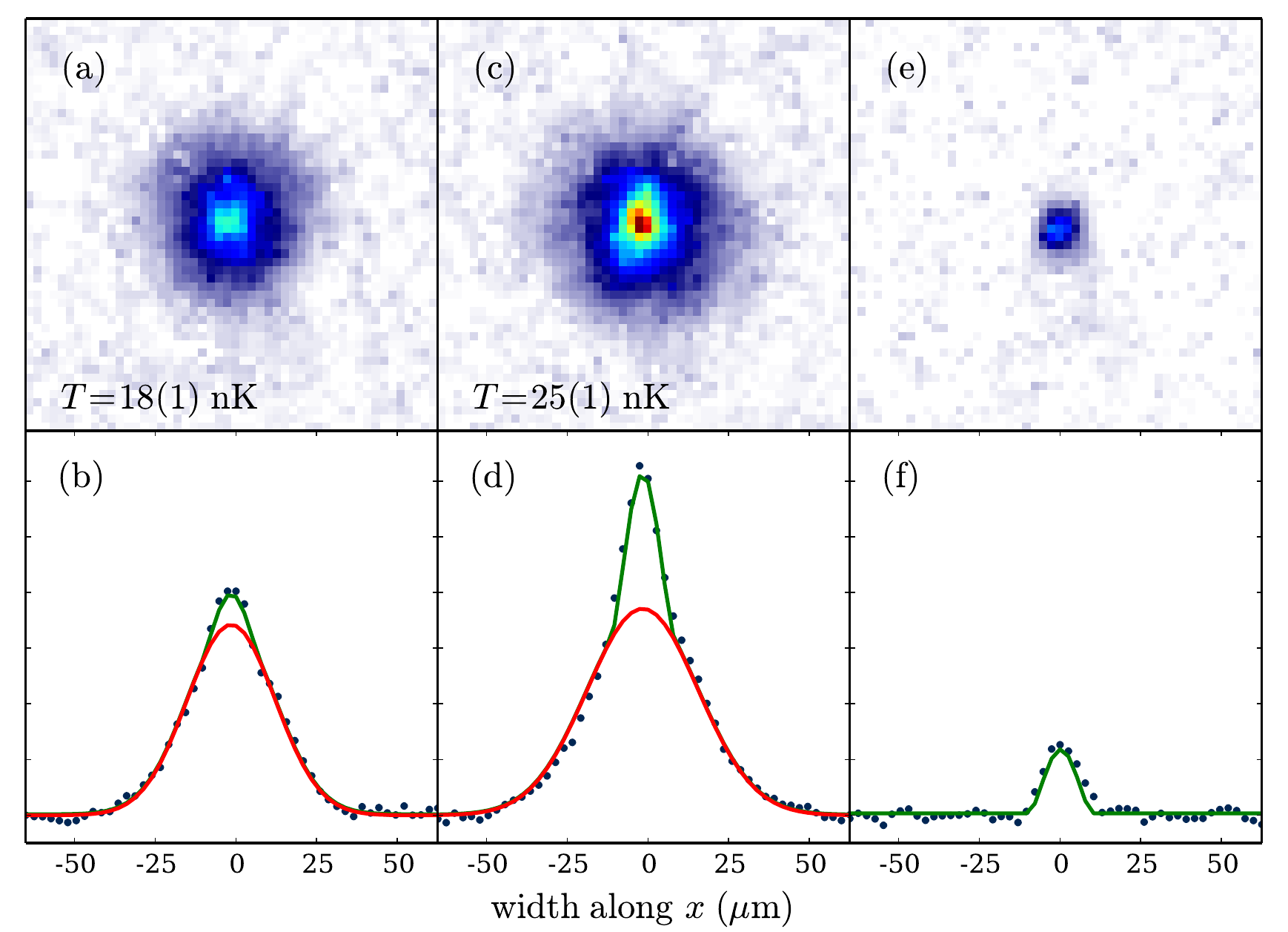}
\caption{Typical single-shot density profiles of $^{160}$Dy atomic clouds after 15 ms of TOF. First row are absorption images and second row are the corresponding integrated density along $\hat{x}$. Dots are data, green curve is fitted bimodal profile and red is the fitted thermal portion. (a) and (b): $7.6(8)\times10^3$ atoms at the onset of BEC. $T=18(1)$ nK and BEC fraction is approximately 5\%. (c) and (d): Intermediate stage of BEC with $1.1(1)\times10^4$ atoms at $T=25(1)$ nK. BEC fraction is approximately 14\%. (e) and (f): Nearly pure BEC of $8.0(8)\times10^2$ $^{160}$Dy atoms.} 
\label{160BEC}
\end{figure}

\begin{figure}[t]
\includegraphics[width=0.8\columnwidth]{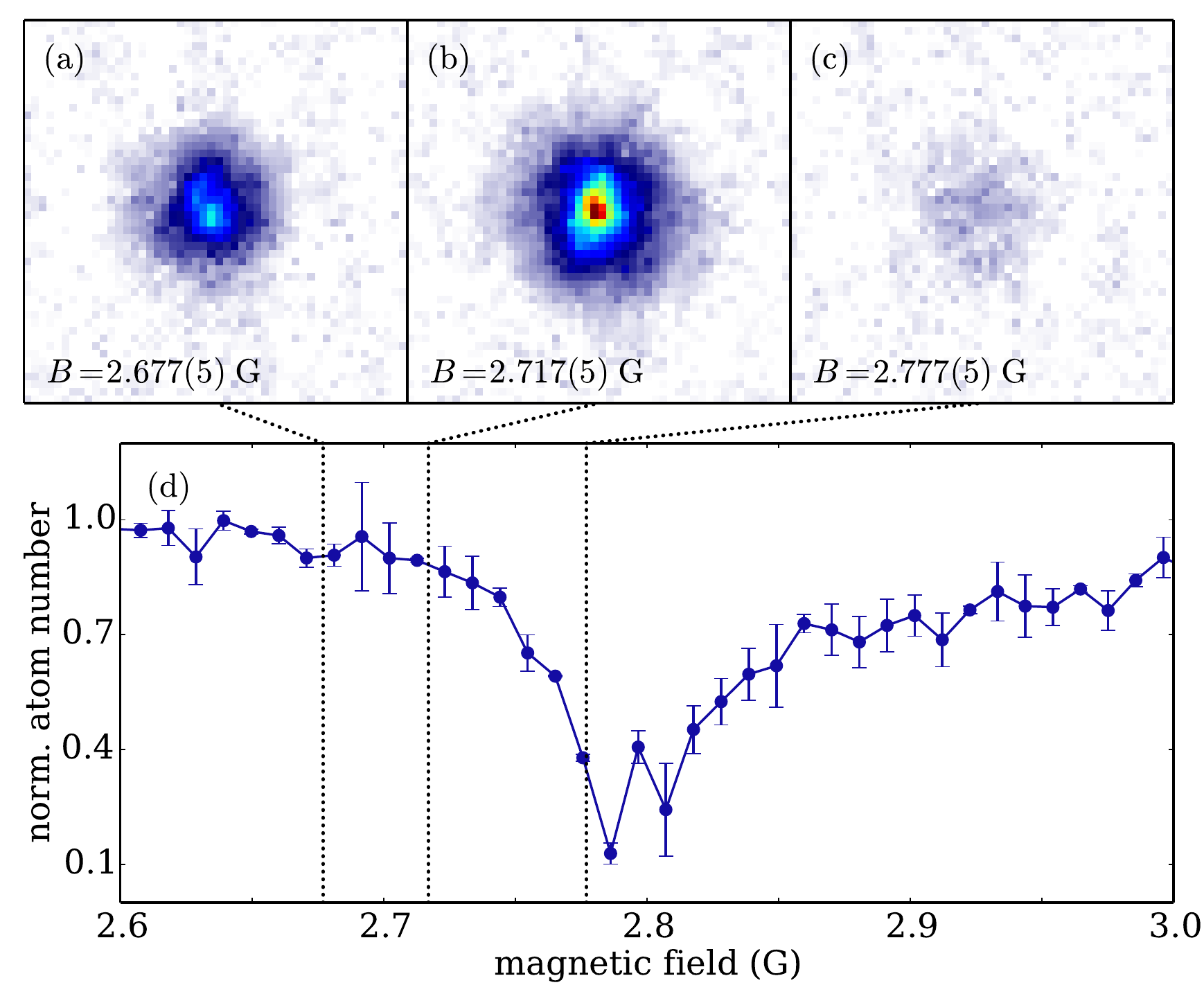}
\caption{Influence of the magnetic field on the production of $^{160}$Dy BECs. Identical evaporative cooling ramps were repeated at three different fields near a Fano-Feshbach resonance at 2.8 G. (a) Away from the resonance, at $B_z=2.677(5)$ G, a cloud of thermal atoms at the threshold of condensation. (b) BEC forms at $B_z=2.717(5)$ G. (c) Rapid atom losses are suffered closer to the resonance near 2.8 G, presumably due to an enhanced rate of inelastic three-body collisions. (d) Atom-loss spectrum of the Fano-Feshbach resonance. The dotted lines mark the value of the magnetic field $B_z$.} 
\label{field}
\end{figure}

A useful method for changing the scattering length of ultracold atoms is to tune the magnetic field near a Fano-Feshbach resonance~\cite{FBChin}. We mapped out a suitable Fano-Feshbach resonance around $B=2.8$~G by performing atom-loss spectroscopy~\cite{FBPRA}, as shown in figure \ref{field}(d). We found it is possible to produce BECs of $^{160}$Dy in at $B_z=2.717(5)$ G on the low-field shoulder of the resonance. Figure \ref{160BEC}(a), (c), and (e) show single-shot absorption images of the cloud after 15 ms of TOF with different levels of condensation achieved. Figure \ref{160BEC}(a) and (b) show the onset of condensation for $7.6(8)\times10^3$ atoms at $T=18(1)$ nK. An intermediate-stage BEC with $1.1(1)\times10^4$ atoms at $T=25(1)$ nK is shown in figure \ref{160BEC}(c) and (d) with a clear bimodal density distribution. Finally, a pure BEC of $8.0(8)\times10^2$ $^{160}$Dy atoms is shown in figure \ref{160BEC} (e) and (f). The thermal portion is not visible, preventing us from determining a temperature. The final powers of ODT2 and ODT3 for the pure BEC are 320(20) mW and 270(10) mW, respectively. The corresponding trap frequencies are [$\omega_x,\omega_y,\omega_z$] = $2\pi\times$[45(3), 23(3), 100(3)] Hz. 

We explored the influence of $B_z$ on $^{160}$Dy BEC production  by repeating the evaporation ramp that produced the BEC in figure \ref{160BEC}(c) at different fields near the Fano-Feshbach resonance. Far away from the resonance, at $B_z=2.677(5)$ G, the evaporation sequence does not provide high-enough efficiency and only a cloud of thermal atoms is achieved, as shown in figure \ref{field}(a). Very close to the resonance, at $B_z=2.777(5)$ G, inelastic loss dominates, presumably due to three-body collisions, and very few atoms survive the evaporation sequence, as shown in figure \ref{field}(c). Evaporating at fields higher than the center of the resonance also does not lead to the formation of a BEC. Using this evaporation sequence, $^{160}$Dy BECs were only produced within a small region of magnetic fields with a width of 40 mG centered around $B_z=2.717(5)$ G.

\section{Conclusions}

We present data demonstrating the Bose-Einstein condensation of two additional isotopes of dysprosium, $^{162}$Dy and $^{160}$Dy. We experimentally found $^{162}$Dy to have favorable scattering properties for evaporative cooling, leading to an order of magnitude increase in condensed atom number compared to the previously published $^{164}$Dy BEC. $^{160}$Dy, on the other hand, is harder to condense, but condensation can be achieved with the use of a Fano-Feshbach resonance. Detailed measurements of these bosonic Dy isotopes'  scattering lengths will be reported~\cite{BohnWithUs}. The large atom number of the $^{162}$Dy BEC makes it an excellent system for studying quantum many-body physics.

\ack{This work is supported by the AFOSR and NSF. YT acknowledges support from the Stanford Graduate Fellowship.}

\section*{References}

\providecommand{\newblock}{}

\end{document}